\documentclass[twocolumn,aps,showpacs,prb,tightenlines,amsmath,amssymb]{revtex4}
\usepackage{bm}
\usepackage{graphicx}
\usepackage{amssymb}               
\usepackage{amsmath}
\usepackage{colordvi} 
\newcommand{\bgreek}[1]{\mbox{\boldmath$#1$\unboldmath}}
\begin{document} 

\title{Electron spin diffusion at the interface of multiferroic oxides}
 
\author{P. Zhang}
\author{M. W. Wu}
\thanks{Author to whom correspondence should be addressed}
\email{mwwu@ustc.edu.cn.}
\affiliation{Hefei National Laboratory for Physical Sciences at
  Microscale and Department of Physics, 
University of Science and Technology of China, Hefei,
  Anhui, 230026, China}
\date{\today}

\begin{abstract} 
We study the spin diffusion in a two-dimensional electron gas
 at the interface of oxide heterostructure 
LaAlO$_3$/SrTiO$_3$ grown on multiferroic TbMnO$_3$ at 15~K by means
of the kinetic spin Bloch equation  
approach. The spiral magnetic moments of Mn$^{3+}$ in TbMnO$_3$
interact with the diffusing spins at the LaAlO$_3$/SrTiO$_3$
interface via the Heisenberg 
exchange interaction. It is demonstrated that the 
spin diffusion length is always finite, 
despite the polarization direction of the injected spins. Our study
also reveals the important role played by the Coulomb scattering,
which can effectively suppress the spin diffusion.

\end{abstract}
\maketitle
\section{Introduction}
During the past decades, spin-based electronics has been a growing area of research due to its
promising applications.\cite{Awschalom,Zutic,Dyakonov,wuReview,fabian565} Among
various electron systems, the two-dimensional electron gas (2DEG) has been widely studied
in spin injection, spin relaxation, spin transport and spin Hall
effect, etc..\cite{ohno790,jonker8180,ohishi2007,wuReview,fabian565,Tombros_07,wunderlich047204,kato306,valenzuela,castro,tapash,cundiff,sherman,tobias1} 
Most of the 2DEG
systems are achieved in the semiconductor quantum wells or
heterostructures.\cite{book} 
The strict 2DEG, e.g., the one existing in a graphene monolayer, 
also attracts much attention 
recently.\cite{ohishi2007,Tombros_07,castro,Geim_07,tapash,yzhou,zhanggraphene} 
Besides, the 2DEG has also been
experimentally realized at the interface of insulating oxides such as LaAlO$_3$/SrTiO$_3$,\cite{thiel1942,ohtomo423,huijben556} opening
the way for the oxide-based nanoelectronics. Starting from this
oxide heterostructure, very recently Jia and Berakdar proposed a 
trilayer system
for new functionalities, which is grown as LaAlO$_3$/SrTiO$_3$/TbMnO$_3$ along
the $c$-axis of orthorhombic 
TbMnO$_3$.\cite{jia014432,jia012105,jiaarxiv1012.4865,jia045309} As a
multiferroic material, TbMnO$_3$ can exhibit both magnetic and
electric orders simultaneously and provide a unique opportunity to
exploit the multifunctionality of a single material.\cite{kenzelmann087206,gajek296,hur392}
SrTiO$_3$ in the middle of the trilayer system is a few layers thick
so that the 2DEG at the interface of LaAlO$_3$/SrTiO$_3$ can be
affected by the magnetic order in TbMnO$_3$. Based on this structure,
Jia and Berakdar predicted the appearance of a persistent spin current
in the 2DEG due to the spiral geometry of the 
local magnetic order.\cite{jia014432} They also proposed a flash
memory model based on the claim that the
injected spins in the 2DEG have no decay of spin 
polarization along the ${a}$-axis when travelling along the
${b}$-axis.\cite{jia012105} This claim was
inferred\cite{jia012105} from the fact that the 
spin polarization along the [110]
direction does not decay in (001) GaAs quantum-well-based 2DEG
with identical Rashba\cite{rashba} and Dresselhaus\cite{dresselhaus} 
spin-orbit coupling strengths, in the {\em absence} of any external magnetic
field.\cite{cheng205328,averkiev271,schliemann146801,tobias} 
Nevertheless, this claim is questionable in the 2DEG formed at
the interface of the oxide, because 
the local magnetic moments
serve as a static magnetic field in the Voigt configuration in the spiral
frame\cite{jia012105} and, as pointed out by Weng and Wu,\cite{weng235109}
a magnetic field in the Voigt configuration can effectively
suppress spin diffusion/transport even when the D'yakonov-Perel' 
spin-orbit coupling\cite{dp} is 
absent.\cite{weng235109,zhang075303} 
In this work, we investigate the spin diffusion
along the ${b}$-axis at the interface of
LaAlO$_3$/SrTiO$_3$ closely above TbMnO$_3$ by means of the kinetic
spin Bloch equation (KSBE) approach.\cite{wuReview} It is found
that the spin diffusion length is {\em finite} in spite of the spin 
polarization direction and the Coulomb scattering makes marked effect on the
spin relaxation in diffusion .

This paper is organized as follows. In Sec.~\ref{haks}, we give the
Hamiltonian of the 2DEG in the collinear frame with 
${\bf\hat {\bgreek \sigma}}_z$ parallel to the $c$-axis, and in the spiral
frame with ${\bf\hat {\bgreek \sigma}}_z$ parallel to the local
magnetic moment, respectively. We then present the KSBEs in the spiral
frame. In Sec.~\ref{sd} we study the spin
diffusion based on the KSBEs. We
summarize in Sec.~\ref{conclusion}.

\section{Hamiltonian and KSBEs}
\label{haks}

The 2DEG formed at the LaAlO$_3$/SrTiO$_3$ interface of
LaAlO$_3$/SrTiO$_3$/TbMnO$_3$ compounds is schematically shown in
Ref.~\onlinecite{jia014432}. We assume it is confined by an
  infinitely deep square potential well with width $a=3$~nm.\cite{thiel1942} The coordinate system is set as
${\bf \hat x}={\bf \hat b}$, ${\bf \hat y}=-{\bf \hat a}$ and
${\bf \hat z}={\bf \hat c}$. When the
temperature $T$ is lower than the ferroelectric Curie temperature $T_c$
($\sim$ 27~K), Mn$^{3+}$ magnetic moments in TbMnO$_3$ form a spiral order with the local magnetic
moment ${\bf M}({\bf r})=M_x\sin({\bf  q}\cdot{\bf r}){\bf\hat x}+M_z\cos({\bf q}\cdot{\bf r}){\bf\hat
  z}$, where ${\bf q}=(0.27\times\frac{2\pi}{b},0,\frac{2\pi}{c})$ is the
modulation vector.\cite{kenzelmann087206,yamasaki147204,malashevich037210,arima097202} $b=0.586$~nm 
and $c=0.749$~nm are the lattice parameters of
TbMnO$_3$.\cite{kenzelmann087206} In reality $M_x\neq M_z$ (e.g., $M_x/M_z\approx 1.4$ at
$T=15$~K),\cite{yamasaki147204,arima097202,kenzelmann087206,malashevich037210} however for simplicity we
take $M_x\approx M_z=M$ following Jia and Berakdar.\cite{jia014432,jia012105} It is assumed
that the 2DEG only interacts with the local magnetic momentums of
Mn$^{3+}$ on the surface of TbMnO$_3$ (i.e., the plane $z=0$) via the Heisenberg
exchange interaction.\cite{jia014432,jia012105} As a
result, the Hamiltonian of the 2DEG reads\cite{jia014432,jia012105}
\begin{equation}
H=\frac{{\bf P}^2}{2m^\ast}+J{\bf\hat n}_{\bf r}\cdot{\bgreek\sigma},
\label{hamiltonian}
\end{equation}
with the first term on the right-hand side of the equation representing the kinetic energy
and the second term, the exchange interaction. $m^\ast$ is the
effective electron mass set to be 10$m_e$ ($m_e$ is the free-electron
mass),\cite{jia014432,jia012105} $J$ stands for the coupling strength, ${\bf \hat n}_{\bf
  r}=(\sin(q_xx),0,\cos(q_xx))$ denotes the unit vector along the local
magnetic moment located on $z=0$ plane, ${\bf P}$ is the momentum operater and 
${\bgreek\sigma}$ is the vector of the Pauli matrices. Performing 
a local rotation around the ${y}$-axis in the spin space as ${\tilde H}=U_g^\dagger(x) HU_g(x)$ with $U_g(x)=e^{-iq_xx\sigma_y/2}$, one obtains in the spiral frame\cite{jia014432,jia012105}
\begin{equation}
{\tilde H}=\frac{1}{2m^\ast}[(P_x-\frac{\hbar q_x\sigma_y}{2})^2+P_y^2]+J\sigma_z.
\label{rhamiltonian}
\end{equation}
This Hamiltonian can be written in the momentum space as
\begin{equation}
{\tilde H}=\frac{\hbar^2{\bf k}^2}{2m^\ast}+{\bf h_k}\cdot{\bgreek\sigma},
\label{khamiltonian}
\end{equation}
in which
\begin{equation}
{\bf h_k}=(0,\frac{\hbar^2q_x}{2m^\ast}k_x,J)
\label{hk}
\end{equation}
and a uniform energy displacement $\varepsilon_0=\frac{\hbar^2q_x^2}{8m^\ast}$ is omitted. The position-dependent 3$\times$3 orthogonal rotation matrix $R(x)$, which obeys $U_g^\dagger(x){\bgreek \sigma}U_g(x)=R(x){\bgreek \sigma}$, reads
\begin{equation}
R(x)=\left(\begin{array}{ccc}
  \cos (q_xx) & 0 & \sin (q_xx) \\
  0  & 1 & 0 \\
  -\sin (q_xx) & 0 &  \cos (q_xx)
\end{array}\right).
\label{rotation}
\end{equation}
$R^T(x)$ transforms any spin-vector in the spiral frame back
to the collinear frame. Our study is performed by first solving the KSBEs in
the spiral frame and then obtaining the spin diffusion properties in
the collinear frame with the aid of $R^T(x)$.

In the study, polarized spins are injected at $x=0$ and diffuse along
the ${x}$-axis. The system is uniform along the fixed
${y}$-axis in both frames. In the spiral frame, the KSBEs  read\cite{wuReview}
\begin{eqnarray}\nonumber
 && \frac{\partial\rho_{\bf k}(x,t)}{\partial
    t}+\frac{e}{\hslash}\frac{\partial \Psi(x,t)}{\partial
  x}\frac{\partial \rho_{{\bf k}}(x,t)}{\partial
  k_{x}}+\frac{\hbar k_x}{m^\ast}\frac{\partial \rho_{\bf
      k}(x,t)}{\partial x}\\ &&\mbox{}+\frac{i}{\hbar}[{\bf h_k}\cdot{\bgreek\sigma},
\rho_{\bf k}(x,t)] 
+\frac{\partial\rho_{\bf k}(x,t)}{\partial t}\Big|_{\mbox{scat}}=0,
\label{KSBE}
\end{eqnarray}
where $\rho_{\bf k}(x,t)$ are the single-particle density matrices of electrons
with wave-vector ${\bf k}$ at position $x$ and time $t$. $\frac{e}{\hslash}\frac{\partial \Psi(x,t)}{\partial
  x}\frac{\partial \rho_{{\bf k}}(x,t)}{\partial
  k_{x}}$ are the driving terms with the electric potential 
satisfying the Poisson equation $\nabla^2_x\Psi(x,t)=e[N_e(x,t)-N_0]/(a
\kappa_0 \varepsilon_0)$, where $N_e(x,t)=\sum_{\bf k}$Tr$[\rho_{\bf
  k}(x,t)]$ stands for the electron density at position $x$ and time
$t$, $N_0$ is the background positive charge
density and $\kappa_0\approx 24$ is the relative static dielectric
constant.\cite{konaka283,kozuka487} 
In the absence of the external electric field, the driving
term is negligible. $\frac{\hbar k_x}{m^\ast}\frac{\partial \rho_{\bf
    k}(x,t)}{\partial x}$,  $\frac{i}{\hbar}[{\bf h_k}\cdot{\bgreek\sigma},
\rho_{\bf k}(x,t)]$ and $\frac{\partial\rho_{\bf k}(x,t)}{\partial
  t}\Big|_{\mbox{scat}}$ in Eq.~(\ref{KSBE}) are the diffusion, coherent and scattering
terms, respectively. We lack detailed information of the electron-phonon and 
electron-magnon scatterings in this material, except that 
 the electron-phonon scattering in SrTiO$_3$ was 
suggested to be typically very weak.\cite{huijben556}
 However, as we 
are interested only in the low temperature behavior (we take
$T=15$~K$<T_c$ in this paper),
both scatterings are neglected. The electron-impurity scattering is also
neglected in the high-mobility 2DEG (for electron density $N_e\sim
10^{13}$~cm$^{-2}$
and mobility $\sim 10^4$~cm$^2$/(V$\cdot$s),\cite{thiel1942} the impurity density
is calculated to be $\sim 10^8$~cm$^{-2}$, corresponding to which the
electron-impurity scattering is found to be negligible). The
scattering left in our consideration is the 
electron-electron Coulomb scattering, which has been proved to be
important in spin relaxation and spin transport in
semiconductors.\cite{wu373,wuReview,cheng083704,weng075312} The
detailed expression of the Coulomb scattering can be found in
Ref.~\onlinecite{weng245320}. By solving the KSBEs one obtains the
spin diffusion properties from $\rho_{\bf k}(x,+\infty)$, the steady-state distribution of density matrices.

\section{spin diffusion}
\label{sd}
\subsection{Spin relaxation in time domain}
Equation~(\ref{hk}) is analogous to the spin-orbit coupling in 
(001) GaAs quantum-well-based 2DEG with identical Rashba and 
Dresselhaus spin-orbit coupling strengths (only the linear Dresselhaus term 
is considered) when $J$ is set to zero. There,
when the ${x}$-axis is set along [1${\bar 1}$0], 
the spin-orbit coupling  reads ${\tilde {\bf h}}_{\bf k}
=(0, 2\gamma k_x, 0)$, with $\gamma$ 
being the coefficient of the Rashba (or Dresselhaus) spin-orbit 
coupling.\cite{schliemann146801,averkiev15582,cheng083704,tobias} 
Therefore, if the spin polarization is along the ${y}$-axis, there
is no spin precession and hence  the spin relaxation is 
suppressed.\cite{schliemann146801,averkiev15582,cheng083704,tobias} 
However, the exchange interaction in oxide provides
 a static magnetic field $J$ along the ${z}$-axis in the spiral frame, 
as shown in Eq.~(\ref{hk}). This field rotates the spin polarization 
away from the ${y}$-axis. The spin polarization away from the ${y}$-axis
 feels the inhomogeneous broadening,\cite{wuReview,wu373} i.e., 
the momentum-dependent spin-orbit coupling from the ${y}$-component
 of ${\bf h_k}$, which, together with any scattering,
 leads to irreversible spin 
relaxation.\cite{wu373,cheng083704,wuReview} Moreover, the
effective magnetic field mixes spin relaxations along the ${x}$-${z}$
plane to the one along the ${y}$-axis, similar to the case based on semiconductor quantum wells.\cite{cheng083704,tobias}
Therefore, the spin relaxation time along the ${y}$-axis is 
not infinite as claimed by Jia and Berakdar,\cite{jia012105}
neither does the spin relaxation along other directions. 
The above scenario also happens to (110) GaAs quantum wells. 
  The effective magnetic field induced by the Dresselhaus spin-orbit
  coupling is oriented along the quantum-well growth direction when
only the lowest subband is relevant, which leads to an infinite spin
relaxation time when the spin polarization is along that direction.
However, by applying a static magnetic field in the
 quantum-well plane, the spin relaxation time 
becomes finite.\cite{wu509,dohrmann147405}

\begin{figure}[thb]
{\includegraphics[width=8cm]{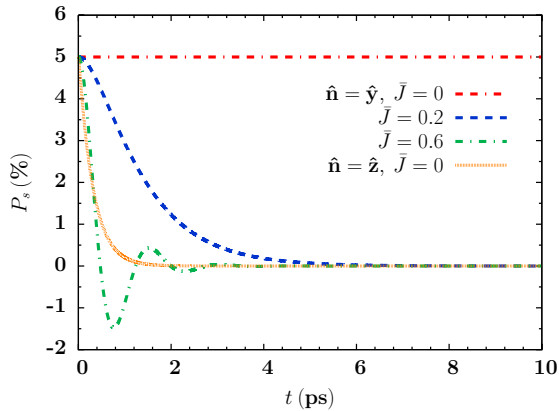}}
    \caption{(Color online) Time evolution of spin polarization along
      different directions under different $J$ (in the figure ${\bar
        J}=\frac{m^\ast}{\pi N_e\hbar^2}J$).} 
   \label{figzw1}
 \end{figure}

 To be quantitative, we
numerically solve the KSBEs in time 
domain\cite{wuReview,cheng083704,weng245320,weng235109,weng075312}
with the Coulomb scattering explicitly included. In Fig.~\ref{figzw1}, the time
evolution of spin polarization (with an initial value 5~\%) is
plotted. It is clearly shown that when the initial spin 
polarization ${\bf \hat n}$ is along ${\bf\hat y}$, the spin
polarization relaxes when $J\neq 0$. However, for other spin
polarization directions such as ${\bf\hat z}$, spin
relaxes even when $J=0$.

\subsection{Spin diffusion: analytical and numerical study}
In this section we investigate the spin diffusion of the 2DEG. We
first consider a simplified case without the scattering for which the
KSBEs can be  solved analytically. Then we numerically solve the KSBEs
in the presence of the Coulomb scattering.

\begin{figure}[thb]
 {\includegraphics[width=8cm]{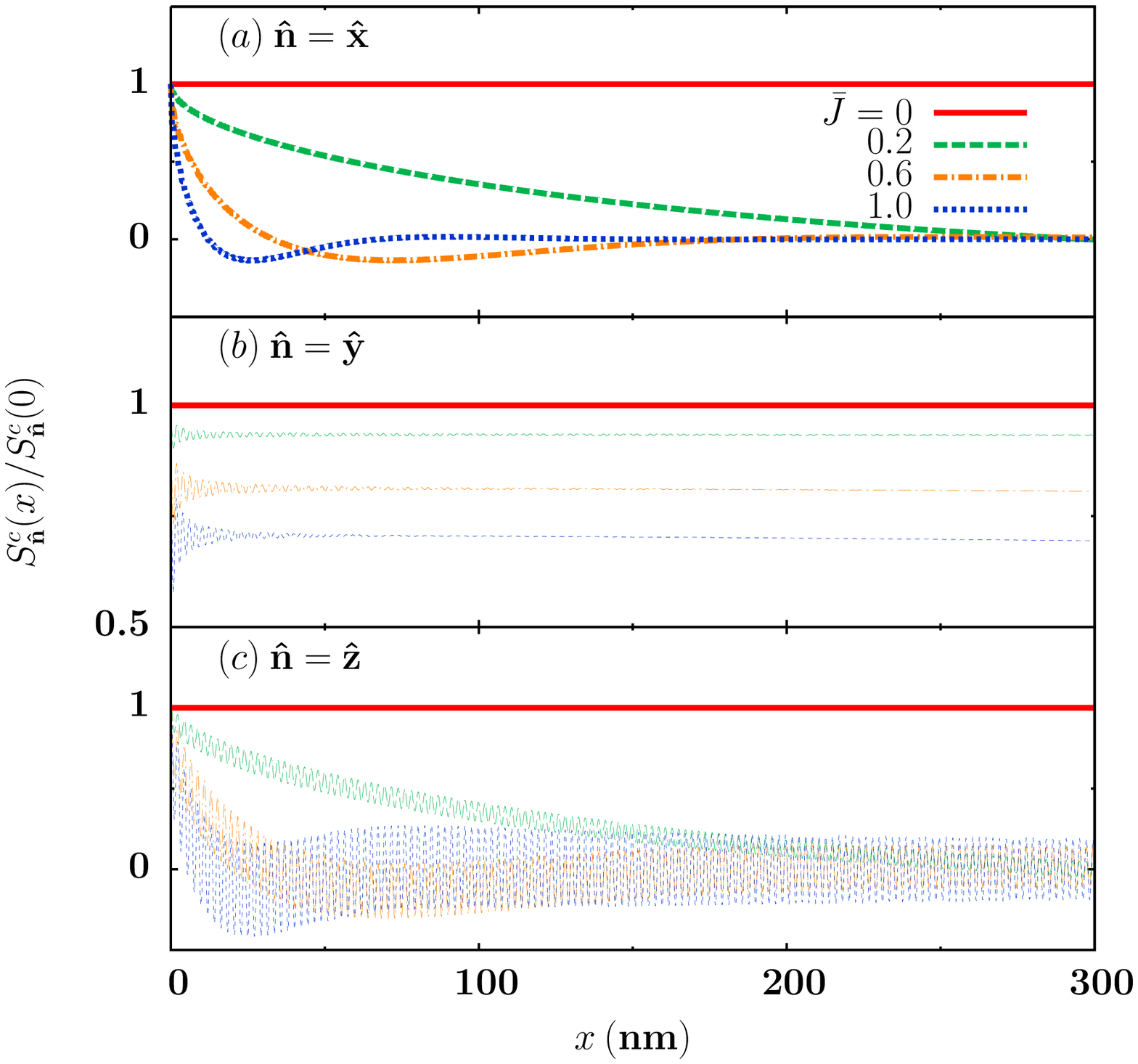}}\\
 {\includegraphics[width=8cm]{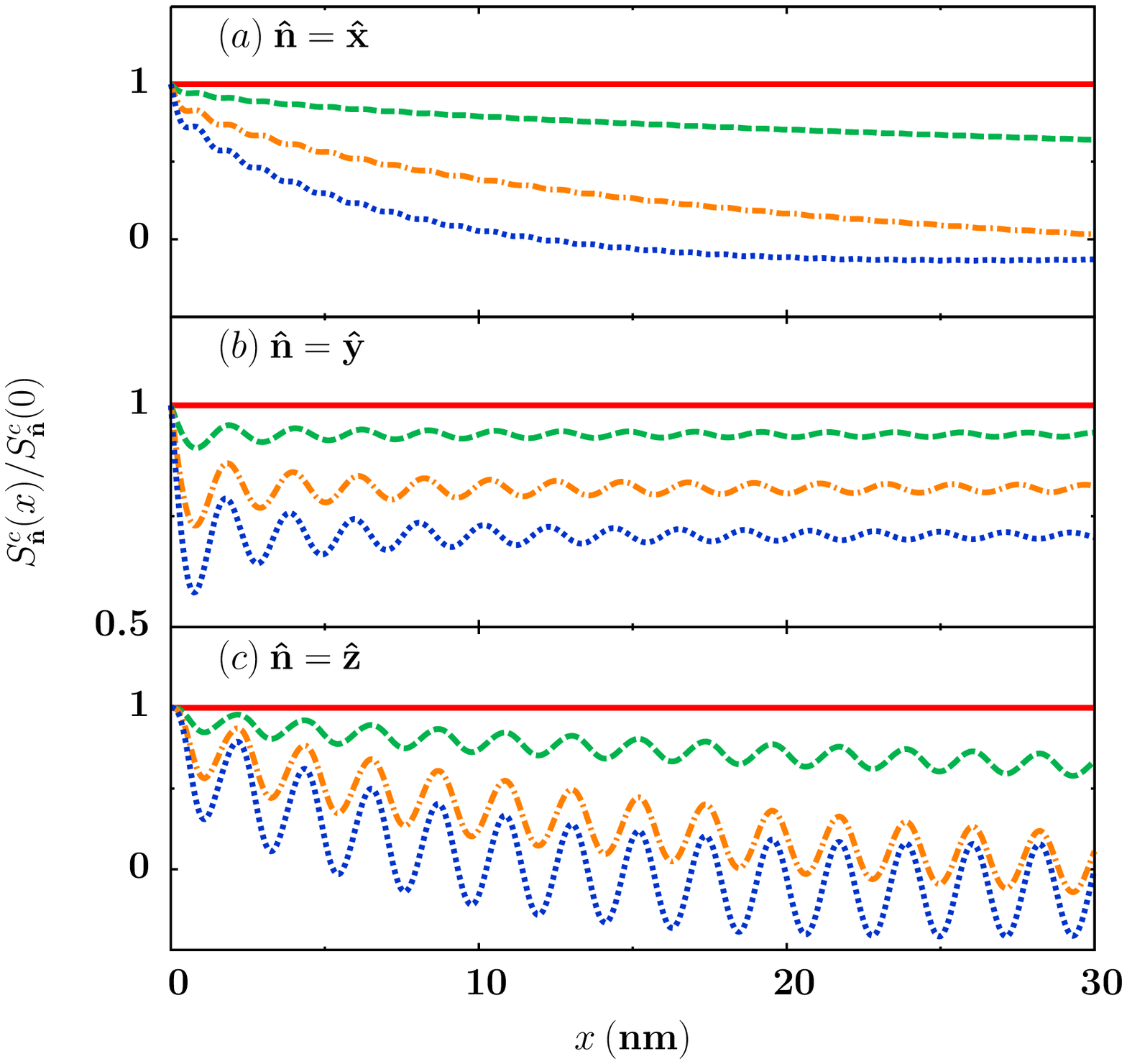}}
    \caption{(Color online) $S^c_{\bf\hat n}(x)/S^c_{\bf\hat n}(0)$ {\em vs}. $x$ for
different ${\bf\hat n}$ and ${\bar J}$. (a) and (d): ${\bf\hat n}={\bf\hat x}$;
(b) and (e): ${\bf\hat n}={\bf\hat y}$; (c) and (f): ${\bf\hat n}={\bf\hat
  z}$. (a)-(c) are plotted in a large  scale of $x$ while (d)-(f) are in
a small scale.}
   \label{figzw2}
 \end{figure} 

When the scattering terms are excluded (and the driving terms are
absent), the KSBEs can be solved in the 
 steady state as 
\begin{equation}
  \rho_{\bf k}(x,+\infty)=e^{-i{\bgreek \omega_{\bf k}}\cdot{\bgreek
      \sigma}x/2}\rho_{\bf k}(0,+\infty)e^{i{\bgreek \omega_{\bf
        k}}\cdot{\bgreek\sigma}x/2}
    \label{noscat}
\end{equation}
with 
\begin{equation}
{\bgreek\omega}_{\bf k}=\frac{2m^\ast}{\hbar^2k_x}{\bf  h_k}=(0, q_x, \frac{2m^\ast}{\hbar^2k_x}J)
\label{frequency}
\end{equation}
 depicting the spin precession frequency in spatial
domain. It has been shown that the inhomogeneous broadening in spin
diffusion/transport is determined by ${\bgreek\omega}_{\bf k}$ rather than 
${\bf  h_k}$ in the time domain,\cite{wuReview,weng235109,cheng083704,cheng205328,zhang075303} with $k_x$ arising from the diffusion term. It becomes evident that
$J$, playing the role of a static magnetic field in the time domain, now
leads to  the inhomogeneous broadening due to $k_x^{-1}$ and hence a finite 
spin injection length. Similar effect was first predicted theoretically 
by Weng and Wu back in 2002 (Ref.~\onlinecite{weng235109}) 
and  realized experimentally in bulk
silicon.\cite{appelbaum295,zhang075303} 
Moreover, due to the presence of $q_x$ in Eq.~(\ref{frequency}), the spin
diffusion length is finite even when the injected spins are polarized
along the $z$-axis.

The spatial distribution of spin polarization is obtained from the 
density matrices given by
Eq.~(\ref{noscat}). We consider the states with $k_x>0$ because 
only these states can propagate along the ${x}$-axis and 
the states with $k_x<0$ are not spin-polarized in the absence of
scattering.\cite{cheng083704} 
Therefore the density matrices with
$k_x>0$ at $x=0$ are fixed as boundary conditions. The spin 
polarization  ${\bf S}_{\bf k}(x)\equiv\mbox{Tr}[\rho_{\bf
  k}(x,+\infty){\bgreek \sigma}]$ is calculated to be ${\bf
  S_k}(x)=F_{\bf k}(x){\bf S_k}(0)$, with 
\begin{widetext} 
\begin{equation}
F_{\bf k}(x)=\left(\begin{array}{ccc}
    \cos(\omega_{\bf k}x) & -{\hat h}_{{\bf k},z}\sin(\omega_{\bf
      k}x)
    & {\hat h}_{{\bf k},y}\sin(\omega_{\bf k}x)  \\
    {\hat h}_{{\bf k},z}\sin(\omega_{\bf k}x) & {\hat h}_{{\bf
      k},y}^2+{\hat h}_{{\bf k},z}^2\cos(\omega_{\bf k}x) & {\hat h}_{{\bf
    k},y}{\hat h}_{{\bf k},z}[1-\cos(\omega_{\bf k}x)] \\
-{\hat h}_{{\bf k},y}\sin(\omega_{\bf k}x) & {\hat h}_{{\bf
    k},y}{\hat h}_{{\bf k},z}[1-\cos(\omega_{\bf k}x)] & {\hat h}_{{\bf
      k},z}^2+{\hat h}_{{\bf k},y}^2\cos(\omega_{\bf k}x)
\end{array}\right), 
\end{equation}
\end{widetext}
in which $\omega_{\bf k}=|{\bgreek\omega}_{\bf k}|$, ${\hat h}_{{\bf
    k},y(z)}=h_{{\bf k},y(z)}/|{\bf h_k}|$ and ${\bf S_k}(0)=\mbox{Tr}[\rho_{\bf
  k}(0,0){\bgreek\sigma}]$. The total spin signal in the spiral
frame is 
\begin{equation}
{\bf S}(x)=\sum_{k_x>0}{\bf S_k}(x)=\sum_{k_x>0}F_{\bf k}(x){\bf S_k}(0),
\label{sr}
\end{equation}
and in the collinear frame reads
\begin{equation}
{\bf S}^c(x)=R^T(x){\bf S}(x).
\label{sl}
\end{equation}

\begin{figure}[thb]
 \hspace{0 cm}   {\includegraphics[width=8cm]{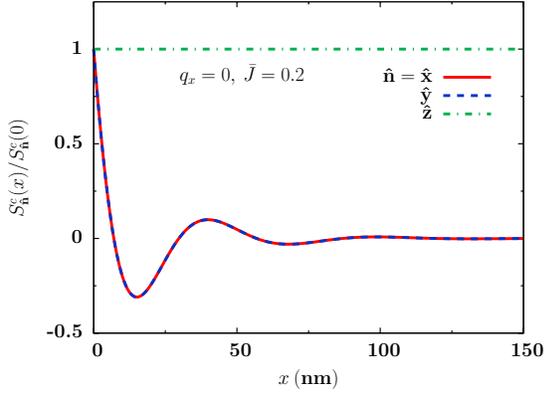}}
 \caption{(Color online)  $S^c_{\bf\hat n}(x)/S^c_{\bf\hat n}(0)$ {\em vs.} $x$ for
different ${\bf\hat n}$ when $q_x$ is set to be zero. ${\bar J}=0.2$.}
   \label{figzw3}
 \end{figure}

In Fig.~\ref{figzw2}, the projection of spin signal
${\bf S}^c(x)$ on the injected spin polarization direction
${\bf\hat n}$, $S^c_{\bf\hat n}(x)\equiv {\bf S}^c(x)\cdot{\bf\hat n}$,
calculated from Eq.~(\ref{sl})
is plotted against $x$ with different $J$ [the data are scaled
  by $S^c_{\bf\hat n}(0)$].
In the computation, we set ${\bf S_k}(0)=[f(\varepsilon_{\bf
  k}-\mu_+)-f(\varepsilon_{\bf  k}-\mu_-)]{\bf\hat 
n}$.\cite{cheng083704,cheng205328,zhang075303}
 $f(\varepsilon_{\bf k}-\mu_\xi)$ with $\varepsilon_{\bf
    k}=\hbar^2k^2/2m^\ast$ is the Fermi distribution under temperature $T$.
The chemical potential $\mu_\xi$ is determined by
$N_e$, the electron density of the 2DEG, and $P_s$, the polarization 
of injected spins, at $x=0$ via the relation $\sum_{\bf k}f(\varepsilon_{\bf
    k}-\mu_\xi)=0.5(1+\xi P_s)N_e$.\cite{cheng083704,cheng205328,zhang075303}
  We set $N_e=10^{13}$~cm$^{-2}$ (Ref.~\onlinecite{thiel1942}) and 
  $P_s$=5~\%.  ${\bf\hat n}$ is  set as ${\bf\hat x}$, 
${\bf\hat y}$ and ${\bf\hat z}$.  It is shown that when $J\neq 0$, the
spin polarization decays during spin diffusion. When $J$ 
increases, the decay becomes faster. It is noted that $S^c_{\bf\hat
   n}(x)$ relaxes to a finite value for ${\bf\hat n}={\bf\hat
   y}$ in the absence of scattering. However, 
when the scattering is included, the
 spin polarization will completely relax to zero 
during spin diffusion
(this will be shown later).
Moreover, the spatial distribution of spin polarization shows
 high-frequency oscillations when $J\neq 0$,
as detailed in Figs.~\ref{figzw2}(d)-(f) in a smaller scale of $x$. 
The frequency of these oscillations is comparable to
the rotation frequency of the spiral magnetic moments $q_x$ and insensitive
to the exchange coupling strength. For
comparison, we recalculate $S^c_{\bf\hat n}(x)$ by replacing the
spiral order with the ferromagnetic order, i.e., 
 aligning all the magnetic  moments 
uniformly along the ${z}$-axis. Hence
 $q_x=0$. The corresponding results are shown in
Fig.~\ref{figzw3}, where the high-frequency oscillations disappear and
the spin polarization along the $z$-axis does not decay due to the
absence of the inhomogeneous broadening. 

\begin{figure}[thb]
 \hspace{0 cm}   {\includegraphics[width=8cm]{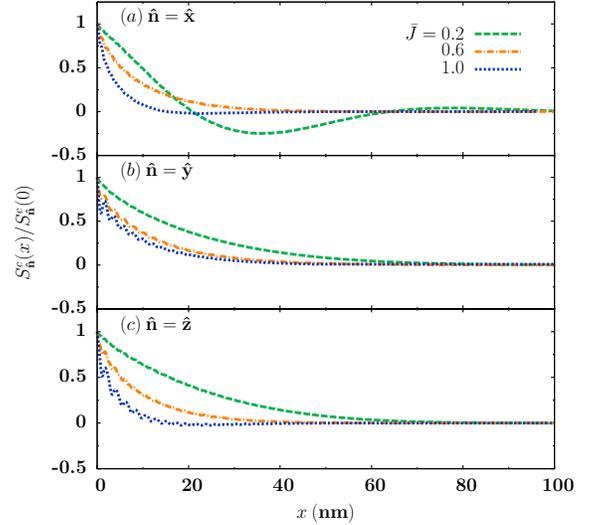}}
 \caption{(Color online)  $S^c_{\bf\hat n}(x)/S^c_{\bf\hat n}(0)$ {\em vs.} $x$ for
different ${\bf\hat n}$ and ${\bar J}$. Here $S^c_{\bf\hat n}(x)$ are
obtained by numerically solving the KSBEs in the presence of the
Coulomb scattering.}
   \label{figzw4}
 \end{figure}

It is noted that the above decay of spin polarization 
along the spin diffusion  in the absence of scattering 
  is due to the interference effect.\cite{weng235109} 
Now we numerically solve the full KSBEs with the Coulomb scattering explicitly
included, facilitated with the double-side injection boundary
conditions.\cite{wuReview,cheng205328,zhang075303,zhanggraphene}
For the cases with $J\neq 0$, we plot $S_{\bf\hat n}^c$ as 
a function of position $x$ in Fig.~\ref{figzw4}. It is shown that the Coulomb
scattering effectively leads to the spin relaxation along the spin diffusion.
 By comparing Figs.~\ref{figzw4} and
\ref{figzw2}, it is observed that with the scattering, the spin
diffusion is suppressed along all directions.\cite{zhang075303} Especially, the spin polarization along the $y$-axis relaxes to zero completely
during the spin diffusion. It is also noted that
the amplitude of the high-frequency oscillations shown in
Fig.~\ref{figzw2} becomes much  smaller.

\section{conclusion}
\label{conclusion}
In conclusion, we have studied the spin diffusion in a 2DEG at the
interface of oxide heterostructure LaAlO$_3$/SrTiO$_3$ closely above the
multiferroic TbMnO$_3$ by means of the KSBE approach. At low temperature 
($< 27$~K), Mn$^{3+}$
magnetic moments in TbMnO$_3$ form a spiral order and couple with the
2DEG via the Heisenberg exchange interaction. It is found that in
 the presence of the spiral magnetic order, in contrast to the claims in the
literature, the spin
polarization decays during spin diffusion, despite the polarization 
direction of injected spins. We also report that the
electron-electron Coulomb scattering suppresses spin diffusion
effectively at  low temperature.

\begin{acknowledgments}
We would like to thank J. Berakdar for valuable discussions.
This work was supported by the National Natural Science
Foundation of China under Grant No. 10725417 and the Knowledge
Innovation Project of Chinese Academy of Sciences. 

\end{acknowledgments}

\end{document}